\documentclass[12pt]{article}
\usepackage[utf8]{inputenc}
\usepackage{amsthm}
\usepackage{amsmath}

\usepackage{amssymb}
\usepackage{caption}
\usepackage{placeins}
\usepackage{subcaption}

\usepackage{enumerate}
 \usepackage[shortlabels]{enumitem}
\usepackage{tikz}
\usepackage{hyperref}
\usepackage[ruled,linesnumbered]{algorithm2e}
\newcommand{\augnetworkflow}{\hat{\mathcal{G}}}
\newcommand{\ehat}{\hat{e}}
\newcommand{\Ehat}{\hat{E}}
\newcommand{\Bijin}{B_{ij}^{\mathrm{in}}}
\newcommand{\Bijout}{B_{ij}^{\mathrm{out}}}
\newcommand{\bijin}{b_{ij}^{\mathrm{in}}}
\newcommand{\bijout}{b_{ij}^{\mathrm{out}}}
\newcommand{\xhat}{\hat{x}}
\newcommand{\flow}{\hat{f}}

\newcommand{\vhat}{\hat{v}}

\newcommand{\Vaug}{\hat{V}}
\newcommand{\Vaugij}{\hat{V}_{ij}}
\newcommand{\bminv}{\underline{b}_v}

\newcommand{\Eijone}{\hat{E}_{ij}}
\newcommand{\Eijtwo}{\hat{E}_{ij,i'j'}}
\newcommand{\Eijconnect}{\hat{E}_{ij,j+1}}

\newcommand{\pathopt}{\path^*}

\newcommand{\auge}{\tilde{e}}

\newcommand{\augc}[1]{w_{#1}}
\newcommand{\augV}{\tilde{V}}
\newcommand{\pathext}{{P}}
\newcommand{\chargeext}{\hat{q}}
\newcommand{\Qext}[1]{\hat{\Q}_{#1}}
\newcommand{\qhat}{\hat{q}}
\newcommand{\qopt}{q^*}
\newcommand{\batteryin}{b^{\mathrm{in}}}
\newcommand{\Batteryin}{B^{\mathrm{in}}}
\newcommand{\Batteryout}{B^{\mathrm{out}}}
\newcommand{\batteryout}{b^{\mathrm{out}}}

\newcommand{\battery}{d}
\renewcommand{\time}{\ell}

\renewcommand{\path}{P}
\newcommand{\charge}{q}

\newcommand{\changebat}{\lambda}
\newcommand{\network}{\mathcal{G}}
\newcommand{\range}{L}
\newcommand{\xmin}{x^{\dagger}}

\newcommand{\minbat}[1]{{\underline{b}_{#1}}}
\newcommand{\maxbat}[1]{{\bar{b}_{#1}}}

\newcommand{\origin}{s}
\newcommand{\dest}{t}
\newcommand{\speed}{r}
\newcommand{\nslot}{a}
\newcommand{\Q}{\mathcal{Q}}
\newcommand{\Qbat}{\Q}
\newcommand{\maxthresh}{J}
\newcommand{\pathend}{m}
\DeclareMathOperator*{\argmax}{arg\,max}
\DeclareMathOperator*{\argmin}{arg\,min}

\usepackage[margin=1in]{geometry}
\newtheorem{theorem}{Theorem}
\newtheorem{lemma}{Lemma}
\newtheorem{proposition}{Proposition}

\newtheorem{assumption}{Assumption}
\theoremstyle{definition}

\newtheorem{example}{Example}

\usepackage{fullpage}
\usepackage{graphicx}
\allowdisplaybreaks
\begin{document}

\title{Network Flow Problems with Electric Vehicles}
\author{%
Haripriya Pulyassary\\
\footnotesize{Cornell University, School of Operations Research and Information Engineering, hp297@cornell.edu}\and Kostas Kollias\\
\footnotesize{Google Research, kostaskollias@google.com}\and Aaron Schild\\
\footnotesize{Google Research, aschild@google.com}\and
David Shmoys\\
\footnotesize{Cornell University, School of Operations Research and Information Engineering, david.shmoys@cornell.edu}\and
Manxi Wu \\
\footnotesize{Cornell University, School of Operations Research and Information Engineering, manxiwu@cornell.edu}
}

%\institute{\and}
%
\date{}
\maketitle              % typeset the header of the contribution
\begin{abstract}
Electric vehicle (EV) adoption in long-distance logistics faces challenges such as range anxiety and uneven distribution of charging stations. Two pivotal questions emerge: How can EVs be efficiently routed in a charging network considering range limits, charging speeds and prices? And, can the existing charging infrastructure sustain the increasing demand for EVs in long-distance logistics? This paper addresses these questions by introducing a novel theoretical and computational framework to study the EV network flow problems. We present an EV network flow model that incorporates range constraints and nonlinear charging rates, and identify conditions under which polynomial-time solutions can be obtained for optimal single EV routing, maximum flow, and minimum-cost flow problems. Our findings provide insights for optimizing EV routing in logistics, ensuring an efficient and sustainable future.
\end{abstract}
%
% We develop efficient computational methods for computing the optimal routing and flow vector using a novel graph augmentation technique.
%
\section{Introduction}

Electric vehicle (EV) adoption for long-distance logistics faces the challenge of range anxiety. Due to the weight of the cargo and constraints on battery sizes, charging can take a considerable proportion of the trip time and requires the use of dedicated charging facilities. Additionally, the uneven distribution and limited availability of charging stations in specific regions or along certain transportation routes introduce further complications for long-distance logistics planning. Two essential questions arise: First, how to efficiently route a single EV or a fleet of EV in a charging network given the battery constraint, charging speed and prices? Second, given the existing charging network, what is the maximum EV flow that can be supported? Answering these questions is crucial for the efficient deployment of EVs in logistics services, and assessing if the current charging infrastructure can support the growing EV demand in long-distance logistics. 

 %Our paper make three contributions: First, we introduce Second, we pinpoint precise conditions under which the optimal single EV routing problem, maximum flow, and minimum-cost flow problems are solvable in polynomial time. Lastly, we introduce a network augmentation method that enables the efficient computation of optimal routing for both single EVs and EV fleets, as well as the maximum flow of the charging infrastructure. 
%We consider a directed graph $\network = (V, E)$, where $V$ is the set of nodes, and $E$ is the set of edges. The network has a set of $K$ origin-destination (o-d) pairs $\{(s_k, t_k)\}_{k \in K}$.  A subset of nodes $I \subseteq V$ are installed with chargers of different types (e.g., DC fast chargers, level-one and level-two chargers). Each node $i \in I$ is equipped with $a_{i} \in \mathbb{N}_{\geq 0}$ chargers; the speed of the charger is a piecewise-linear function of the vehicle's battery level $b$. We route a fleet of electric vehicles through this network, where each vehicle has a battery range limit $L$ and is directed between their origin and destination. The vehicle's routing strategy specifies the chosen path $P$ and the battery charge $q$ acquired at each visited station $i \in P$. A charging strategy is feasible if the battery level $b$ remains above zero, and does not exceed its range limit $L$. 

To address the above-mentioned questions, we develop an EV network flow model that incorporates range constraints and nonlinear charging speeds. We consider a directed graph, in which a subset of nodes are equipped with electric vehicle charging stations. Chargers across different stations can vary in charging speed (e.g., DC fast chargers, level-one and level-two chargers). The charging speed, measured as the percentage of battery replenished over time, depends on the vehicle's current battery level — the higher the battery level, the slower the charging rate. We route a fleet of electric vehicles through this network, where each vehicle has a battery range limit and is directed between their origin and destination. The vehicle's routing strategy specifies the chosen path and the battery charge acquired at each visited station. A charging strategy is feasible if the battery level remains above zero, and does not exceed its range limit. % within its constraints, neither depleting nor exceeding its range limit. 

We study three fundamental problems in this setting:  (i) the \emph{single EV optimal charging strategy} problem that computes a charging strategy that minimizes its total (monetary and time) cost; (ii) the \emph{maximum EV flow} problem that computes the maximum volume of EV fleet that can be served by the charging network given its station capacity constraints; and (iii) the \emph{minimum-cost EV flow} problem that routes an EV fleet with a fixed volume to minimize the total (time and monetary) costs. These problems can be viewed as natural extensions of the shortest path, maximum flow, and minimum-cost flow problems in the classical network flow models, but with two new hurdles: First, given that the feasible charging strategy is a vector with continuous values, the corresponding flow vector, representing the volume of the EV fleet adopting each feasible charging strategy, is infinite-dimensional. Second, as a station's charging speed changes with the battery level of the arriving vehicle, the station's effective capacity inherently depends on the flow vector. Consequently, the EV flow problems studied in this work differ non-trivially from their standard network flow counterparts \cite{ahuja1988network,ford1962flows,goldberg1988new}. %Therefore, new approach is needed to address these problems in EV settings. %To navigate these challenges, we cast the maximum flow problem as a semi-infinite linear program with infinite number of variables and a finite set of constraints (\cite{LOPEZ2007491,romeijn1992duality}). Our investigation reveals that the optimal flow only takes positive value with respect to a finite set of feasible charging strategies, which are the extreme points of the feasible charging strategy set. We introduce an augmented graph, where each path represents an extreme point feasible charging strategy. Therefore, the charging network's maximum flow can be delineated as the capacity of this augmented network and solved using a finite linear program in polynomial time.

%In both of the EV flow problems ((ii) and (iii)), we consider capacity constraints on the amount of charge that can be obtained from a station per unit time. Such constraints cannot be captured by standard edge-capacity constraints, and consequently, the EV flow problems studied in this work differ non-trivially from their standard network flow counterparts. In addition to the complex and highly-coupled nature of the station capacity constraints, another significant technical challenge faced in this setting is the infinite-dimensional nature of charging strategies (as the amount of charge obtained from a station by a vehicle is a continuous quantity). 

We show that the three problems described above are NP-hard in general settings. We pinpoint precise conditions under which we can provide polynomial-time algorithms to solve them.  
In problem (i), we employ a dynamic programming approach to solve the optimal single EV charging strategy under the assumption that the path with the minimum battery consumption between the origin and any other node is also the path with the minimum travel time cost (Assumption \ref{as:proportional}). Our algorithm generalizes the previous studies \cite{khuller_fill_2007,sweda2012finding} by incorporating nonlinear charging speed and general charging cost functions. In problems (ii) and (iii), we cast the flow problems as semi-infinite linear programs with an infinite number of variables and a finite set of constraints (\cite{LOPEZ2007491,romeijn1992duality}). Our investigation reveals that there exists an optimal flow vector that  utilizes only a finite number of feasible charging strategies, which are the extreme points of the feasible charging strategy set. Building on this observation, we construct a charge-augmented network, where each path represents an extreme point feasible charging strategy. When the network  has capacity constraints only on the charging stations, both the maximum flow problem and the minimum-cost flow problem can be solved using a polynomial-sized linear program using the charge-augmented network. We also provide fully polynomial time approximation schemes for solving the network flow problems in the general setting with edge capacity constraints. %all three problems in general settings when the problems become using the equivalence of (approximate) separation and (approximate) optimization. 

Our results also contribute to the rich literature of EV routing and charging that includes path planning and charging of EV fleets in delivery problems \cite{chen2023exponential,desaulniers2016exact,flath2014improving,kullman2021electric,froger2022electric,parmentier2023electric,sweda2017adaptive,kinay2023charging,adler2017vehicle,qi2023scaling}, ride-hailing operations with EVs \cite{dong2022dynamic,kullman2021electric}, infrastructure planning and battery swapping system design \cite{mak2013infrastructure,schneider2014electric,yu2022coordinating,qi2023scaling}. This line of literature primarily focuses on developing computational methods of path generation and charging scheduling. Our focus is different in that we aim at developing efficient network algorithms for long-distance logistics, and assessing the capacity of charging networks.  
 
%The problem of computing an optimal charging strategy for a single EV has been well-studied in the literature, and several FPTAS algorithms have been proposed for this problem in general settings \cite{baum2015shortest,merting2015routing}.  Indeed, our dynamic programming algorithm can be viewed as an extension of the algorithms in \cite{khuller_fill_2007,sweda2012finding} which only considered linear charging functions and did not consider edge-costs. The other two problems we study in this work are EV flow problems, which are similar in spirit to classical network flows . As discussed earlier, there are non-trivial differences between EV flows and classical network flows; in particular, many fundamental results, including the celebrated maximum-flow minimum-cut theorem, do not hold in the EV setting. To the best of our knowledge, ours is the first work to formalize and study this flow model with station capacity constraints. 

\color{black}

\section{The Charging Network Model}\label{sec:model}
We study the problem of routing electric vehicles in a directed graph $\network = (V, E)$, where $V$ is the set of nodes, and $E$ is the set of edges. The network has $K$ origin-destination (o-d) pairs $\{(s_k, t_k)\}_{k \in K}$.
A subset of nodes $I \subseteq V$ are installed with chargers of different types. Each node $i \in I$ is equipped with $a_{i} \in \mathbb{N}_{\geq 0}$ number of chargers. Chargers at different nodes may have different charging speeds, which we will detail shortly.\footnote{When a node is equipped with multiple types of chargers, we can equivalently view each type of chargers as a separate node. These nodes are connected by edges with zero distances.}

Electric vehicles are each equipped with a battery of size $\range$. We assume that all vehicles are fully charged when departing from their origin, and origin and destination nodes $\{s_k, t_k\}_{k \in K}$ have no chargers. As a vehicle traverses an edge $e \in E$, it consumes $\battery_e \geq 0$ amount of battery, and takes time $\time_e \geq 0$. A vehicle that travels between any o-d pair $(s_k, t_k)$ selects a charging strategy $(\path, \charge)$, where $\path : s_k=v_0\, v_1\, \dots\, v_m\, v_{m+1}=t_k$ is an $s_k$-$t_k$ path in the network, and $\charge = (q_v)_{v \in V}$ indicates the amount of battery charged at each node. A charging strategy $(\path, \charge)$ is feasible if it satisfies the following constraints: %the battery level of a vehicle remains to be non-negative and does not exceed $\range$ along the path, and the charge amount can only be positive at stations that are on the path, i.e. 
\begin{subequations}\label{eq:feasCharge}
\begin{align}
\batteryin_{v_i} &:= \sum_{n=0}^{i-1} q_{v_n} + L - \sum_{n=0}^{i} d_{(v_{n-1}, v_{n})} \geq 0, \quad \forall i = 1, \ldots, \pathend + 1, \label{eq:feasCharge1} \\  
\batteryout_{v_i} &:= \sum_{n=0}^i q_{v_n} + L - \sum_{n=0}^{i} d_{(v_{n-1}, v_{n})} \leq L, \quad \forall i = 1, \ldots, \pathend, \label{eq:feasCharge2} \\  
q_{v} &=0, \quad \forall v \in V \setminus \{\path \cap I\}, \qquad q_{v} \geq 0, \quad \forall v \in V, \label{eq:feasCharge3}
\end{align}
\end{subequations}
where $\batteryin_{v_i}$ as in \eqref{eq:feasCharge1} is the battery level of a vehicle when arriving at $v_i$ that is non-negative, and $\batteryout_{v_i}$ as in \eqref{eq:feasCharge2} is the battery level when leaving $v_i$ that does not exceed $L$. The constraint \eqref{eq:feasCharge3} ensures that the charge amount is non-negative, and charge can only happen at charging stations that are on the path $\path$. 
For a given $\origin_k$-$\dest_k$ path $P$, the set of feasible charging vector $q$ is defined as
\begin{equation}
Q_P = \{q \in \mathbb{R}^{|V|} : q \text{ satisfies } \eqref{eq:feasCharge1}-\eqref{eq:feasCharge3}\}.  
\end{equation}
Here, $Q_P$ is a polytope. Let $\mathcal{P}_k$ be the set of all $\origin_k$-$\dest_k$ paths and $\Q_k= \{(P, q) : q \in Q_P, P \in \mathcal{P}_{k}\}$. We define the set of all paths as $\mathcal{P} = \cup_{k \in K} \mathcal{P}_k$ and the set of all feasible charging strategies as $
     \Q  = \left\{ (P, q)\right\}_{q \in Q_P, P \in \mathcal{P}}$.

%As a vehicle traverses a road segment, its battery is consumed is proportional to the length of the road segment. We note that this charge-discharging model, while simplistic, is a reasonable model for road networks that do not have huge variation in elevation. Henceforth, we abuse notation slightly and use ``distance of an edge $e$'' and ``amount of charge required to traverse the edge $e$'' interchangeably, and denote this quantity by $d_e$. 

We next define the charging speed, and the cost of a feasible charging strategy. One key feature of electric vehicle charging is that the  charging speed (i.e., the amount of battery charged per unit of time) decreases as the battery level of the vehicle increases. We assume that the charging speed is a piecewise linear functions of a vehicle's battery level $b \in [0, L]$, and the speed of charger at node $i$ is $\speed_i(b)$ given by: 
\begin{align}\label{eq:charge_rate} \speed_i(b) =  \speed_{ij},  \quad \forall \alpha_j \leq b < \alpha_{j+1}, \quad \forall j \in \{1, \ldots, \maxthresh\},\end{align}
where  $\alpha_1, \ldots, \alpha_{\maxthresh+1}$ are fixed thresholds with $0 = \alpha_1 < \alpha_2 <  \cdots < \alpha_{\maxthresh+1}= L$, and $\speed_{ij}>0$ is the speed at which a vehicle charges if its present battery level is between $\alpha_j$ and $\alpha_{j+1}$ (refered as the $j$-th interval). Here, the thresholds $\{\alpha_j\}_{j=1}^{\maxthresh}$ are set to be the same for all chargers although their charging speeds may differ in each interval. This is without loss of generality since we can define the piecewise linear function using the combined threshold sets of all chargers. %We define the \emph{configuration} of a charging network to be the tuple $(N, I, \speed, \nslot, d, \alpha)$. 

The \emph{cost} of a feasible charging strategy $(P, q)$ includes the \emph{monetary} cost for charging at each station, and the time cost of both traversing edges and charging at stations. In particular, vehicles are charged with price $\tau_{ij}>0$ for a unit of battery obtained from charger at node $i$ given that the vehicle's battery level is in the $j$-th interval, and price $\rho_{i} \geq 0$ for occupying a charging spot at station $i$ for one unit of time. The cost of a charging strategy $(P, q)$ with $P = v_0\,v_1\,\cdots\, v_\pathend \,v_{\pathend + 1}$ is
\begin{align}\label{eq:charge_cost}
    c(P, q) = \sum_{i=0}^{\pathend} \ell_{(v_i,v_{i+1})} + \sum_{i \in I} \sum_{j = 1}^\maxthresh \left(\tau_{ij} + \frac{1 + \rho_i}{\speed_{ij}}\right) q_{ij},
\end{align} 
where $\sum_{i=0}^{\pathend} \ell_{(v_i,v_{i+1})}$ is the total latency cost of traversing the edges along path $\path$, $q_{ij}$ is the amount of battery at node $i$ with battery level between $\alpha_j$ and $\alpha_{j+1}$, and $(\tau_{ij} + \frac{1 + \rho_i}{\speed_{ij}}) q_{ij}$ is the monetary and time cost of charging $q_{ij}$.\footnote{Given a charging strategy $(P, q)$, $ q_{ij} = \min\left(\alpha_{j+1} - (\batteryin_i + \sum_{n=1}^{j} q_{in}),  q_i - \sum_{n=1}^{j} q_{i n}\right) \times \mathbb{I}\left[\alpha_j \leq \batteryin_i + \sum_{n=1}^{j} q_{in} < \alpha_{j+1}\right]$ for all $ i \in I \cap P$ and all $j \in \{1, \ldots \maxthresh\}$, and $\batteryin_i$ is given by \eqref{eq:feasCharge1}. Moreover, $q_{ij}=0$ for all $i \notin P$ and all $j \in \{1, \dots, J\}$. We can check that $\sum_{j=1}^{J} q_{ij} = q_i$. } %, and , $\frac{\rho_i}{\speed_{ij}} q_{ij}$ is the monetary cost of occupying the charger at $i$ with time $\frac{q_{ij}}{\speed_{ij}}$, and $\frac{1}{\speed_{ij}} q_{ij}$ is the time cost needed for charging $q_{ij}$. In \eqref{eq:charge_cost}, both the monetary and time cost of charge are summed across all chargers and all battery intervals. 

A \emph{flow} vector in the charging network is defined as $x = (x_{P, q})_{(P, q) \in \Q}$, where $x_{P, q}$ is the non-negative flow assigned to each feasible charging strategy $(P, q) \in \Q$. Since the charge vector $q$ is a real-valued vector, $x$ has \emph{infinite dimension}. A flow vector $x$ is feasible if the total amount of battery charged at each station $i \in I$ is less than or equal to the amount of charge that can be dispensed per unit time. Here, we need to account for the fact that the charging speed of each charger as in \eqref{eq:charge_rate} is a piecewise linear function of the battery level of the vehicle. Therefore, the amount of battery charge that each charger can provide depends on $x$. We introduce variables $z = (z_{ij})_{i \in V, j=1, \dots, \maxthresh}$, where each $z_{ij} \geq 0$ governs the allocation of charging capacity of the total $a_i$ chargers across the $\maxthresh$ battery intervals. For each $j$, we must ensure that the total amount of charge consumed by flow $x$ at speed $\speed_{ij}$ from each station $i \in I$ does not exceed the allocated amount $\speed_{ij} z_{ij}$, and the allocation vector $z$ is feasible. Therefore, the set of feasible flow $x$ can be characterized by the following constraints: 
\begin{subequations}
   \begin{align}
    \sum_{(P, q) \in \Q} q_{ij} x_{P,q} &\leq \speed_{ij} z_{ij}, \quad  \forall i \in I, \quad \forall j \in \{1, \ldots, \maxthresh\}, \label{eq:feasflow:cap} \\ 
     \sum_{j = 1}^\maxthresh z_{ij} &= \nslot_i, \quad \forall i \in I, \qquad x, z \geq 0. \label{eq:feasflow:zij}
     %\sum_{(P, q) \in \Q: e \in \path} x_{P, q} &\leq u_e, && \forall e \in E, \label{eq:feasflow:edge} 
     % &\currbat_i = L + \sum_{p = 1}^{i-1} q_p - \sum_{p=1}^{i - 1} d_{(v_p, v_{p+1})}, && \forall i \in I, \label{eq:feasflow:bi} \\ 
     % &q_{ij} = \min(\alpha_{j+1} - (\currbat_i + \sum_{p=1}^{j-1} q_{i, p}),  q_i - \sum_{p=1}^{j-1} q_{i, p}) \cdot \mathbb{I}[\alpha_j \leq \currbat_i + \sum_{p=1}^{j-1} q_{i, p} < \alpha_{j+1}],  && \forall i \in I, ~ \forall j = 1, \ldots \maxthresh. \label{eq:feasflow:qij}
\end{align}
\end{subequations}
Our goal is to develop efficient algorithms for the following problems: \\
(i) \emph{Single EV optimal charging problem}: Given a charging network, o-d pair $(s_k, t_k)$, and monetary charging costs $(\tau, \rho)$, compute
    \begin{equation}\tag{Single-OPT}\label{obj:single_opt}
    (P^*, q^*) \in \argmin_{(P, q) \in \Q_k} ~ c(P, q),\end{equation} 
    where $c(P, q)$ is given by \eqref{eq:charge_cost}. \\
    (ii) \emph{Maximum EV flow problem}: Given a charging network, compute \begin{equation}\tag{$\mathrm{P_{max-flow}}$}\label{P:max}
    x^* \in \argmax_{x}  \sum_{(P, q) \in \Q} x_{P, q}, \quad s.t. \quad \text{$x$ satisfies \eqref{eq:feasflow:cap}-\eqref{eq:feasflow:zij}.} \end{equation} \\
    (iii) \emph{Minimum-cost EV flow problem}: Given a charging network and monetary charging costs $(\tau, \rho)$, $\xmin$ minimizes the total cost of the flow with demand vector $(D_k)_{k \in K}$:
    \begin{equation}\tag{$\mathrm{P_{min-cost}}$}\label{P:mincost}
        \begin{split}
        \xmin \in \argmin_{x} C(x):= \sum_{(P, q) \in \Q} c(P, q) \cdot x_{P,q},\\
        s.t. \quad \text{$x$ satisfies 
 \eqref{eq:feasflow:cap}-\eqref{eq:feasflow:zij}, ~ and $\sum_{(P, q) \in \Q} x_{P, q} \geq D_k, \quad \forall k \in K$}.
    \end{split} 
    \end{equation}

%    In Sections \ref{sec:gasStation} and \ref{sec:maxFlow}, we solve (i) single EV optimal routing problem and (iii) min-cost flow problem under the following assumption: 

% \noindent\emph{1. Single EV optimal charging strategy $(P^*, q^*)$}: Given a charging network and monetary charging costs $(\tau, \rho)$, 
%     \[(P^*, q^*) \in \argmin_{(P, q) \in \Q} ~ c(P, q),\] 
%     where $c(P, q)$ is given by \eqref{eq:charge_cost}. 

%  \smallskip 
%  \noindent\emph{2. Maximum flow vector $\xmax$}: Given a charging network, \[x^* \in \argmax_{x}  \sum_{(P, q) \in \Q} x_{P, q}, \quad s.t. \quad \text{$x$ satisfies \eqref{eq:feasflow:cap}-\eqref{eq:feasflow:zij_nonneg}.} \] 
% \smallskip
% \noindent\emph{3. minimum-cost flow vector $\xmin$}: Given a charging network and monetary charging costs $(\tau, \rho)$, $\xmin$ minimizes the total cost of the flow with total demand $D$, i.e. 
%     \[\xmin \in \argmin_{x} C(x):= \sum_{(P, q) \in \Q} c(P, q) \cdot x_{P,q}, \quad s.t. \quad \text{$x$ satisfies 
%  \eqref{eq:feasflow:cap}-\eqref{eq:feasflow:zij_nonneg}, ~ and $\sum_{(P, q) \in \Q} x_{P, q} =D$}. \]%satisfies station capacity constraints,  $\sum_{(P, q) \in \Q} x_{P, q} \geq D$ and $c(x)= \sum_{(P, q) \in \Q} c(P, q) \cdot x_{P,q}$ is minimized. 

\section{Optimal charging strategy for single EV }\label{sec:gasStation} 
With general battery consumption vector $d=(d_e)_{e \in E}$ and time cost vector $\ell=(\ell_e)_{e \in E}$, (i) single EV optimal charging problem is NP-hard since the \emph{Resource Constrained Shortest Path Problem} is a special case when the network has no charging station \cite{johnson1981np}. In this section, we show that under Assumption \ref{as:proportional}, the optimal single EV optimal routing strategy can be solved in polynomial time using a dynamic programming-based approach.
 \begin{assumption}\label{as:proportional}
For any $i, i' \in I$, $\mathcal{P}_{ii'}$ is the set of paths that connect $i$ and $i'$, and $\argmin_{P_{ii'}\in \mathcal{P}_{ii'}} \sum_{e \in P_{ii'}} d_{e}= \argmin_{P_{ii'}\in \mathcal{P}_{ii'}} \sum_{e \in P_{ii'}} \ell_{e}$. 
 \end{assumption}Assumption \ref{as:proportional} implies that between any pair of nodes, the path with the minimum battery consumption is also the path with the minimum time cost. In practice, this assumption holds when the changes in elevation and speed are small in the network. Given a charging network, we can verify Assumption \ref{as:proportional} in $O(|I|^3)$ time by running the Dijkstra's algorithm to compute the shortest path between each pair of nodes using the battery consumption vector $d$ and the time cost vector $\ell$, respectively. 
 %When it does not hold, the single EV optimal charging problem and the minimum-cost EV flow problem are NP-hard (as the constrained shortest path problem is a special case). In such scenarios, we no longer have an exact polynomial-time algorithm for these problems; however,  we are able to obtain an FPTAS for the minimum-cost EV flow problem (using the FPTAS for the single EV optimal charging strategy problem by \cite{merting2015routing} as a subroutine). 

We construct an auxiliary network with node set $\{(i, j)\}_{i \in I, j \in \{1, \dots, J\}} \cup \{s_k, t_k\}_{k \in K} $, where each $(i, j)$ is the $j$-th copy of the original station $i$ that has $a_i$ number of chargers with charging rate $r_{ij}$. 
For any two stations $i, i' \in I$, we compute the path with the minimum battery consumption (which also has the minimum time cost given Assumption \ref{as:proportional}), and denote $d_{ii'}$ (resp. $\ell_{ii'}$) as the battery consumption (resp. time cost) of that path. In the auxiliary network, we connect $(i, j)$ and $(i', j')$ for any $i, i' \in I$ such that $d_{ii'} \leq L$ and any $j, j' \in \{1, \dots, J\}$. The battery consumption of such an edge $e' =((i, j), (i', j'))$ is $d_{e'} = d_{ii'}$ and the time cost is $\ell_{e'} = \ell_{ii'}$. We also connect $(i, j)$ and $(i, j')$ for any $j, j' \in \{1, \dots, J\}$ and any $i \in I$ with time cost $\ell_{e'}=0$ and battery consumption $d_{e'}=0$. In the auxiliary network, the total number of nodes is $|I| \maxthresh$, and the number of edges is $O(|I|^2\maxthresh^2)$. For each node $(i, j)$, we denote the lower bound of battery level of charging at $(i, j)$ as $\underline{b}_{ij} = \alpha_{j}$ and the upper bound as $\bar{b}_{ij} = \alpha_{j+1}$. 

For any $(s_k, t_k)$, a charging strategy $(P, q)$ is analogously defined as $\path= s_k\, (i_1, j_1)\, \dots\, (i_m, j_m)\, t_k$ being a path between the origin and destination and $q = (q_{ij})_{i \in I, j \in \{1, \dots, J\}}$. A charging strategy $(P, q)$ is feasible if the battery level of arriving at any $(i, j) \in P$ is higher or equal to $\underline{b}_{ij}$ and less than $\bar{b}_{ij}$. Since the auxiliary network connects any pair of nodes that are within the vehicle's range constraint, we can restrict our attention to consider only charging strategies $(\path, \charge)$, where $\charge_{ij}>0$ for any $(i,j) \in \path$. This is without loss of generality since any charging strategy $(\path', q')$ that violates this constraint can be improved by another strategy $(\path, q)$ such that $\path$ removes any nodes with zero charge amount, and the charge amount for the remaining nodes do not change.  %Additionally, the charge cost of the auxiliary node $v'= (i,j)$ is $\tau_{v'} = \tau_{ij}$ and $\rho_{v'} = \rho_i$ for any $j =1, \dots, \maxthresh$ and any $i \in I$. 

To compute $(\pathopt, \qopt)$, we solve the following recurrence, where %define the function $A: V \times [0, L] \to \mathbb{R}$ such that 
$A(i,j, b)$ is the minimum-cost of traveling from a node $(i,j)$ to the destination $\dest_k$, given that the vehicle arrived at $(i, j)$ with $b \in [0, L]$ battery level, and $N(i, j)$ is the set of neighbours of $(i,j)$.  Then,
\clearpage
{\small \begin{align}\label{eq:recurrence}
    &A(i,j, b) = \notag\\
    &\begin{cases} \min\limits_{(i',j') \in N(i,j)} \min\limits_{x \geq 0}  A(i', j', ~b+x - d_{(ij, i'j')}) + c_{ij} x + \ell_{(ij, i'j')}, & \underline{b}_{ij} \leq b \leq \bar{b}_{ij}, \\ \infty, & \text{otherwise}, \end{cases} 
\end{align}}where $ c_{ij}=  \tau_{ij} + \frac{1 + \rho_{i}}{r_{ij}}$ for all $i \in I, j \in \{1, \dots, J\}$ is the per-unit charging cost at $(i, j)$ as in \eqref{eq:charge_cost}. 
%Following the previous section, we work with the augmented network configuration $(N', I', \speed', \nslot', d', \alpha', \minbat{ }, \maxbat{ })$. Recall that for each station $i$ in the original configuration, we now have $\maxthresh$ copies of station $i$; the $j$th copy $i^{(j)}$ corresponds to vehicles charging at station $i$ while having a battery level between $\alpha_j$ and $\alpha_{j+1}$, and this definition is enforced using battery-level constraints.
%For ease of exposition, define the total per-unit charging cost of any $i \in I$ to be $c_i = \tau_{i} + \frac{1}{\speed_i}$.  

%We wish to find a feasible charging strategy $(P, q)$ that satisfies battery-level constraints at each station $i \in I$, and minimizes $\sum_{i \in I} c_i q_i + \sum_{e \in P} \ell_e$. 

 Since the battery level $b$ of a vehicle is a continuous quantity, solving the dynamic program as in \eqref{eq:recurrence} requires discretization of battery levels. Our next lemma shows that in an optimal solution with path $\path$, a vehicle charges at a node $(i_n,j_n)$ either to reach the maximum allowable battery level $\bar{b}_{i_nj_n}$ or to just be able to reach the next node $(i_{n+1}, j_{n+1})$. 

% to the  the set of all possible battery levels for a vehicle arriving at any station is finite, and the size of . %This lemma is a generalization of Lemma 1 in \cite{khuller_fill_2007}. 

\begin{lemma}\label{lem:polyChargeLevels}
   Suppose that $(P^*, q^*)$ is an optimal charging strategy with $P^* = s_k\, (i_1, j_1)\, \cdots\, (i_m, j_m)\, t_k$. Then, $\qopt$ satisfies
   \begin{align}\label{eq:charge_opt_condition}
   \qopt_{i_n, j_n} = \left\{
   \begin{array}{ll}
       \minbat{i_{n+1}j_{n+1}} + d_{(i_nj_n, i_{n+1}j_{n+1})} - \batteryin_{i_nj_n}  &  \quad \text{if $c_{i_nj_n} > c_{i_{n+1}j_{n+1}}$,}\\[2pt]
       \maxbat{i_nj_n} - \batteryin_{i_nj_n}, & \quad \text{otherwise.}
   \end{array}
   \right.
   \end{align}
   % where 
   % \[\batteryin_{v_i} = \sum_{n=1}^{i-1} \qopt_{v_n} + L - \sum_{n=1}^{i-1}d_{v_i, v_{i+1}}, \quad \forall i=1, \dots, m\]
   % is the battery level when vehicle arrive at node $v_i$ given $\qopt$. 
    % \begin{enumerate}
    %     \item If $c_{i_j} > c_{i_{j+1}}$, 
    %     charge $q_{i_j} = \minbat{i_{j+1}} + d_{i_j, i_{j+1}} - \currbat $ units at station $i_j$.  
    %     \item Otherwise, $q_{i_j} = \maxbat{i_j} - \currbat$. 
    % \end{enumerate} 
   Consequently, given any optimal charging strategy, the battery level $\batteryin_{ij}$ of a vehicle arriving at any $(i,j)$ can only take a finite number of values given by 
    \begin{align}\label{eq:Batteryin}
        \Batteryin_{ij} = \{\minbat{ij}\} \cup \{\maxbat{i'j'} - d_{(ij, i'j')} | (i', j') \in N(i,j),  \minbat{ij} \leq \maxbat{i'j'} - d_{(ij,i'j')} \leq \maxbat{ij}\}.
    \end{align} %where $N(i)$ is the set of vertices that are adjacent to $i$, and satisfy $d_{ij} + \minbat{j} \leq \maxbat{i} $. 
\end{lemma}

Thanks to Lemma \ref{lem:polyChargeLevels}, we only need to analyze the value of $A(i, j, b)$ for $b \in \Batteryin_{ij}$ instead of the continuous battery level $b \in [0, L]$. We define an augmented network with the nodes set $\augV := \{(i, j, b)\}_{i \in I, j =1, \dots, J, b \in \Batteryin_{ij}}$. A node $(i, j, b)$ is connected to $(i',j', \underline{b}_{i'j'})$ if $(i',j') \in N(i, j)$ and $c_{i'j'} < c_{ij}$. The cost of such edge $\auge = \left((i, j, b), (i', j', \minbat{i'j'})\right)$ is $\augc{\auge} = (d_{(ij, i'j')} + \minbat{i'j'} - b) c_{ij} + \ell_{(ij, i'j')}$. On the other hand, if $c_{i'j'} \geq c_{ij}$, then node $(i, j, b)$ is connected to $(i', j', \maxbat{ij}-d_{(ij, i'j')})$ with cost $\augc{\auge} = (\maxbat{ij}-b)c_{ij}+ \ell_{(ij, i'j')}$. Following from \eqref{eq:recurrence}, the value of $A(i, j, b)$ for any $(i, j)$ and any $b \in \Batteryin_{ij}$ equals to the cost of the shortest path (with respect to $w$) from the node $(i, j,b)$ to node $(t_k, 0)$, and the value of $A(i, j, b)=\infty$ if such a path does not exist. Moreover, the minimum-cost charging strategy of a single EV with o-d pair $(s_k, t_k)$ corresponds to the shortest $(s_k, L)$-$(t_k, 0)$ path of this augmented network, and thus can be computed by the Dijkstra's algorithm. 
\begin{theorem}\label{thm:gasStation}
    A single EV optimal charging strategy $(P^*, q^*)$ can be computed in $O((|I|J)^6\log(|I|J))$ time.
    \end{theorem}

\section{Maximum and Minimum-cost EV Flow Problems}\label{sec:maxFlow}

In Sec. \ref{subsec:inf_capacity}, we show that the maximum EV flow problem can be computed in polynomial time, and the minimum-cost EV flow problem can be computed in polynomial time under Assumption \ref{as:proportional}. In Sec. \ref{subsec:general}, we generalize our setting to incorporate capacity constraints on edges. We demonstrate that the flow problems become NP-hard, and we provide fully polynomial time approximation schemes. 

%show that the max flow problem is polynomial time solvable if the edge capacity is infinite (Assumption \ref{as:infinite}), and the min-cost flow problem is polynomial time solvable under Assumptions \ref{as:proportional} and \ref{as:infinite}. In Sec. \ref{subsec:general}, we show that when these assumptions are violated, the maxi flow and the min-cost flow can be computed in FPTAS. 

% \begin{assumption}\label{as:infinite}
% For any $e \in E$, $u_e = \infty$. 
% \end{assumption}

% \manxi{If we have space, add one sentence to justify this assumption in logistics setting.}

\subsection{Unconstrained edge capacity}\label{subsec:inf_capacity}
%Under Assumption \ref{as:infinite}, we drop the edge capacity constraint  from \eqref{P:max}. 
%In this section, we present polynomial-time algorithms for the maximum EV flow and minimum-cost EV flow problems. % computing the  \emph{maximum flow vector $\xmax$} and the \emph{minimum-cost flow vector $\xmin$}. 
% In the sequel, we work with the auxiliary network after metric completion. In this network, the maximum EV flow problem can be formulated as the following LP \begin{subequations}
%      \begin{align}
%          \max_{x,z}  ~ &\sum_{(P, q) \in \Q} x_{P, q} \\
%          \text{s.t.} ~& \sum_{(P, q) \in \Q} q_{ij} x_{P,q} \leq \speed_{ij} z_{ij} && \forall (i,j)\in V, \label{eq:feasflow:cap:aug} \\ 
%      ~~&~~ \sum_{j=1}^{J} z_{ij} = \nslot_i && \forall i \in I,  \label{eq:feasflow:zij:aug}\\
%     ~~&~~ z_{ij} \geq 0 && \forall (i,j) \in V,\label{eq:feasflow:zij_nonneg:aug}
%      \end{align}
%  \end{subequations}
%  where $\Q = \cup_{k \in K}\Q_{s_kt_k}$, and $\Q_{st}$ (as defined by \eqref{eq:Qst}) is the set of all feasible charging strategies (in the auxiliary network) for the o-d pair $(s, t)$.  In \eqref{P:max}, constraint \eqref{eq:feasflow:cap:aug} ensures that the total amount of battery charge at station $i$ and $j$-th battery level satisfies the capacity constraint $\speed_{ij}$ multiplied with the capacity allocation $z_{ij}$. Additionally, \eqref{eq:feasflow:zij:aug} and \eqref{eq:feasflow:zij_nonneg:aug} ensures that $(z_{ij})_{j}$ is a feasible allocation of charging slots at each station $i$. 
The dual program of \eqref{P:max} is as follows: %, show that it can be solved efficiently, and use this to construct a polynomial-sized subset of feasible strategies $\Q' \subseteq \Q$ such that an optimal solution to \eqref{P:max} can be obtained by solving the (polynomial-sized) restriction on $\Q'$. The dual of \eqref{P:max} is  
 \begin{subequations}
\begin{align} \min &~ \sum_{i \in I} \nslot_i y_i,  \tag{$\mathrm{D_{max-flow}}$}\label{D:max}  \\
    \text{s.t.}&~  \sum_{i \in I} \sum_{j=1}^{\maxthresh} \pi_{ij} q_{ij} \geq 1, \qquad \forall  (P, q) \in \Qbat,\label{constr:dualsep} \\
    & y_i \geq \pi_{ij} \speed_{ij}, \qquad \qquad \forall i \in I, \quad \forall j \in \{ 1, \ldots, \maxthresh\}, \\ 
    & ~ \pi \geq 0,
\end{align}
\end{subequations}
 where $\pi_{ij}$ and $y_i$ are the  dual variables associated with the primal constraints \eqref{eq:feasflow:cap} and \eqref{eq:feasflow:zij}, respectively.

Since there are an infinite number of feasible charging strategies, and a finite number of stations, both the primal and the dual programs are semi-infinite linear programs. %: the primal \eqref{P:max} has an infinite number of variables $x_{P, q}$ and finite number of constraints since there are infinite number of feasible charging strategies $(P, q) \in \Q$. On the other hand, the dual program \eqref{D:max} has an infinite number of constraints and a finite number of variables. 
For each $P \in \mathcal{P}$, let $\hat{Q}_P$ denote the extreme points of the polytope $ Q_P$, and $\Qext{} = \cup_{P \in \mathcal{P}} \{(P, \hat q) : \hat q \in \hat{Q}_P\} $. The next lemma shows that, since the set of feasible charging vectors is the union of polytopes, we can reformulate the dual program as a finite linear program, where constraints \eqref{constr:dualsep}  need only to be verified for all $(P, \chargeext) \in \Qext{}$. As a result, the primal program is equivalent to a finite linear program. 
\begin{lemma}\label{lemma:finite}
    The dual \eqref{D:max} is equivalent to the following linear program: 
    \begin{subequations}
\begin{align} \min &~ \sum_{i \in I} \nslot_i y_i,  \tag{$\mathrm{D'_{max-flow}}$}\label{D:max_finite}  \\
    \text{s.t.}&~  \sum_{i \in I} \sum_{j=1}^{\maxthresh} \pi_{ij} \hat{q}_{ij} \geq 1, \qquad \forall  (P, \chargeext) \in \Qext{},\label{constr:dualsep:finite} \\
    & y_i \geq \pi_{ij} \speed_{ij}, \qquad \qquad \forall i \in I, ~ \forall j \in \{1, \ldots, \maxthresh\}, \\ 
    & ~ \pi \geq 0.
\end{align}
\end{subequations}
% where $\Qext{st}$ is the set of extreme points of $\Q_{st}$, i.e. 
% \begin{align}
%     \Qext{st} =\left\{(\hat{q}_{ij})_{i \in I, j=1, \dots, \maxthresh} \left\vert \text{\eqref{eq:feasCharge1:aug} or \eqref{eq:feasCharge2:aug} is tight or $\qhat_{ij}=0$, ~ $\forall i \in I, ~\forall j =1, \dots, \maxthresh$}\right.\right\}.
% \end{align}
Additionally, \eqref{P:max} is equivalent to the following linear program: 
 \begin{subequations}
     \begin{align}
         \max_{x,z}  \quad &\sum_{(P, \qhat) \in \Qext{}} \xhat_{\pathext, \qhat}, \tag{$\mathrm{P'_{max-flow}}$}\label{P:max_finite}\\
        s.t. \quad  &\sum_{(P, \qhat) \in \Qext{}} \qhat_{ij} \xhat_{\pathext,\qhat} \leq \speed_{ij} z_{ij}, && \forall i \in I, ~ \forall j \in \{ 1, \ldots, \maxthresh\},  \label{eq:feasflow:cap:aug:finite} \\ 
    & \sum_{j=1}^{\maxthresh} z_{ij} = \nslot_i, && \forall i \in I,  \label{eq:feasflow:zij:aug:finite}\\
    & z_{ij} \geq 0, && \forall i \in I, ~ \forall j \in \{1, \ldots, \maxthresh\}. \label{eq:feasflow:zij_nonneg:aug:finite}
     \end{align}
 \end{subequations}
% where $\xhat =(\xhat)_{(\pathext, \qhat) \in \Qext{}}$. 
\end{lemma}
Lemma \ref{lemma:finite} indicates that the optimal flow can be achieved only by using charging strategies that are in $\Qext{}$. %sending flows of vehicles to follow charging strategies that are in $\Qext{}$. 
% \textcolor{red}{Notably, for any such $\hat q \in Q_P$, \eqref{eq:feasflow:cap} or \eqref{eq:feasflow:zij} is tight or $\qhat_{ij}=0$, ~ $\forall i \in I, ~\forall j =1, \dots, \maxthresh$. }\manxi{better connection needed. I messed up the equation references here.}
We know from Lemma \ref{lem:polyChargeLevels} that for such charging strategies, the set of possible battery levels of vehicles arriving at each node $(i,j)$ has to be in the (polynomial-sized) set $\Batteryin_{ij}$ as in \eqref{eq:Batteryin}. Moreover, a vehicle departing from $(i,j)$ will either have the maximum possible battery level, $\bar{b}_{ij}$,  or the minimum amount required to reach an adjacent node $(i',j')\in N(i, j)$. As a result, the set of possible battery levels of a vehicle leaving node $(i, j)$ is given by: 
\begin{align}\label{eq:Batteryout}
 \Batteryout_{ij} = \{\bar{b}_{ij}\} \cup \{\underline{b}_{i'j'} + d_{(ij, i'j')} | (i', j') \in N(i, j), \,  \underline{b}_{ij} \leq \underline{b}_{i'j'} + d_{(ij, i'j')} \leq \bar{b}_{ij}\}. %~ \forall i \in I, \forall j =1, \dots, J. 
\end{align}

A key insight of the above discussion is that there exists an optimal solution in which every vehicle's charge level upon arrival and departure from a node belongs to a discrete, polynomial-sized set. Due to this observation, we are able to construct a polynomial-sized charge-augmented network  $\augnetworkflow$. For every $i \in I$ and $j \in \{1, \dots, J\}$, we create $|\Bijin| + |\Bijout|$ copies of the node $(i, j)$, denoted as \[\Vaugij= \{(i, j, b) : b \in \Bijin \cup \Bijout\}, \] 
where each node $(i, j, b) \in \Vaugij$ represents a vehicle with battery level $b$ at station-copy $(i, j)$. We define the node set as $\hat{V} = \cup_{i \in I}\cup _{j=1}^{J} \Vaugij$. Moreover, 
%The 
%number of nodes $\Vaugij= \Vaugijin \cup \Vaugijout$, where $\Vaugijin =  \{(i,j,\bijin)\}_{\bijin \in \Bijin}$ represents that the vehicles arrive at station $i$ interval $j$ with battery level $\bijin$, and $\Vaugijout =  \{(i,j,\bijout)\}_{\bijout \in \Bijout}$ represents that the vehicles leave station $i$ interval $j$ with battery level $\bijout$. The set of all nodes in the augmented network is $\Vaug= \cup_{i \in I} \cup_{j=1}^{J} \{\Vaugijin \cup \Vaugijout\}$. 
$\augnetworkflow$ has three types of edges: \begin{itemize}
    \item[--] \emph{Type I (Charging) edges}: $\Eijone : = \{((i, j, \bijin), (i, j, \bijout))| \bijin \in \Bijin, \bijout \in \Bijout, \bijout > \bijin\}$ represents that a vehicle gets a $\bijout - \bijin$ amount of charge from station $i$ and battery interval $j$.\\
    \item[--] \emph{Type II (Station-copy connection) edges}: $\Eijconnect : = \{((i, j, \overline{b}_{ij}), (i, j + 1, \underline{b}_{i,j+1}))\}$ connects consecutive station copies.\\
    \item[--] \emph{Type III edges}: $\Eijtwo: = \{(i, j, \bijout), (i', j', b_{i'j'}^{\mathrm{in}}) | \bijout \in \Batteryout_{ij}, \, b_{i'j'}^{\mathrm{in}} \in \Batteryin_{i'j'}, \, \\ \bijout - b_{i'j'}^{\mathrm{in}} = d_{ij, i'j'}\}$ represents that a vehicle leaves a node $(i, j)$ and moves to another node $(i', j')$ with battery consumption of $\bijout - b_{i'j'}^{\mathrm{in}}$ being equal to $d_{ij, i'j'}$. 
\end{itemize}
Note that a vehicle does not leave the physical station by taking type I and II edges, and hence $d_{\ehat} = \ell_{\ehat} = 0$ for such edges. On the other hand, the battery consumption of a vehicle taking a type III edge $\ehat = ((i, j, b), (i', j', b'))$, is $d_{\ehat}= d_{ii'}$ (and the time cost is $\ell_{\ehat}= \ell_{ii'}$). 

\begin{example}
Consider the network in Figure \ref{fig:orig_network}, with two stations, $i_1$ and $i_2$. All vehicles have battery capacity $L=9$. The charging speed for station $i_1$ is 2 if the vehicle's battery level is in the range $[0, 5]$ and 1 otherwise. For station $i_2$, the charging speed is 3 if the vehicle's battery level is in the range $[0, 5]$ and 2 otherwise. The charge-augmented network is given in Figure \ref{fig:charge_network}. The red arcs correspond to the type I edges for each station-copy. The gray dashed arcs are type II edges connecting consecutive copies of a given station. The black arcs are type III edges. The charging speeds of the green, blue, and yellow nodes are 3, 2, and 1 respectively. 
\begin{figure}[ht]
    \centering
    \begin{subfigure}[b]{0.2\textwidth}
\centering 
    \begin{tikzpicture}[
		roundnode/.style={circle, draw=black!70,fill=white!70, thick, minimum size=1mm},
		squarednode/.style={rectangle, draw=black!60, fill=white!5,  thick, rounded corners, minimum size=1mm},
		]
		%Nodes
		 \draw (0, 0) node[roundnode]      (s)  {$s$};
  	 \draw (1.25, 1) node[squarednode]      (i)  {$i_1$};
    \draw (1.25, -1) node[squarednode]      (j)  {$i_2$};
	\draw (2.5, 0) node[roundnode]      (t)  {$t$};

		%Lines
\draw[->] (s) -- (i) node [midway, above, fill=none]{5};
\draw[->] (s) -- (j) node [midway, below, fill=none]{4};
\draw[->] (i) -- (j) node [midway, right, fill=none]{6};
\draw[->] (j) -- (t) node [midway, below, fill=none]{6};
\draw[->] (i) -- (t) node [midway, above, fill=none]{5};
	\end{tikzpicture}
 \vspace{15pt}
 \caption{}\label{fig:orig_network}
\end{subfigure} 
\begin{subfigure}[b]{0.77\textwidth}
\centering 
\begin{tikzpicture}[
		roundnode/.style={circle, draw=black!70,fill=white!70, minimum size=1mm},
		squarednode/.style={rectangle, draw=black!60, fill=white!5,  thick, rounded corners, minimum size=7mm},
		]
		%Nodes
	\draw (1.5, 0) node[roundnode]      (s)  {$(s, 9)$};
  	\draw (3, 3) node[squarednode, fill=blue!10]      (i15)  {$(i_1, 1, 5)$};
    \draw (3, 2) node[squarednode, fill=blue!10]      (i14)  {$(i_1, 1, 4)$};
    \draw (3, 1) node[squarednode, fill=blue!10]      (i10)  {$(i_1, 1, 0)$};
    \draw (7, 3) node[squarednode, fill=yellow!10]      (i210)  {$(i_1, 2, 9)$};
    \draw (7, 2) node[squarednode, fill=yellow!10]      (i26)  {$(i_1, 2, 6)$};
    \draw (7, 1) node[squarednode, fill=yellow!10]      (i25)  {$(i_1, 2, 5)$};
    \draw (3, -1) node[squarednode, fill=green!10]      (j15)  {$(i_2, 1, 5)$};
    \draw (3, -2) node[squarednode, fill=green!10]      (j13)  {$(i_2, 1, 3)$};
    \draw (3, -3) node[squarednode, fill=green!10]      (j10)  {$(i_2, 1, 0)$};
    \draw (7, -1) node[squarednode, fill=blue!10]      (j210)  {$(i_2, 2, 9)$};
    \draw (7, -2) node[squarednode, fill=blue!10]      (j26)  {$(i_2, 2, 6)$};
    \draw (7, -3) node[squarednode, fill=blue!10]      (j25)  {$(i_2, 2, 5)$};
 \draw (8, 0) node[roundnode]      (t)  {$(t, 0)$};

		%Lines

% s edges
\path[] (s) edge [thick, out=0, in=180, looseness=0]   node[] {} (4, -0.15) 
(4, -0.15) edge [thick, ->, out=0, in=160, looseness=1] node[near end, above] {4} (j25);
\path[] (s) edge [thick, ->, out=90, in=180, looseness=0.5] node[near end, left=2pt] {5} (i14);
\path[] (s) edge [thick, ->, out=-90, in=180, looseness=0.5] node[midway, above=1pt] {4} (j15);
\path[] (s) edge [thick, out=0, in=180, looseness=0]   node[] {} (5, 0.15) 
(5, 0.15) edge [thick, ->, out=0, in=180, looseness=1] node[near end, below] {5} (i25);

% i1 edges 
\path[] (i15) edge [thick, ->, out = 0, in=180, looseness=0, color=gray, dashed] node[near end, below] {} (i25);
\path[] (i15) edge [thick, out = 0, in=180, looseness=.5] (6, .3) 
(6, .3) edge [thick, ->, out = 0, in=180, looseness=0] node[near end, above] {5} (t);

% j1 edges 
%\path[] (j15) edge [thick, out = 0, in=180, looseness=.5]  node[near end, below=1pt] {5} (t);
\path[] (j15) edge [thick, ->, out = 0, in=180, looseness=.5, dashed, color=gray]  node[near end, left] {} (j25);

% i2-t edges 
\path[] (i25) edge [thick, ->, out = 0, in=90, looseness=.5]  node[near end, right] {5} (t);

% j2-t edges 
\path[] (j26) edge [thick, ->, out = 0, in=-90, looseness=1, ]  node[near end, right] {6} (t);

% i->j edges:

\path[] (i26) edge [thick, ->, in = 0, out=180, looseness=.5]  node[near end, below left=7pt] {6} (j10);
\path[] (i210) edge [thick, ->, in = 0, out=180, looseness=.5]  node[near end, below left=7pt] {6} (j13);

% j->i edges
% \path[] (j210) edge [thick, ->, in = 0, out=180, looseness=.5]  node[near end, left] {6} (i14);
% \path[] (j26) edge [thick, ->, in = 0, out=200, looseness=.5]  node[near end, left] {6} (i10);

% Charging edges
\path[] (i10) edge [thick, ->, in = 180, out=180, looseness=1, color=red]  node[near end, below left=5pt] {} (i14)
(i14) edge [thick, ->, in = 180, out=180, looseness=1, color=red]  node[near end, below left=5pt] {} (i15);

\path[] (j10) edge [thick, ->, in = 180, out=180, looseness=1, color=red]  node[near end, below left=5pt] {} (j13)
(j13) edge [thick, ->, in = 180, out=180, looseness=1, color=red]  node[near end, below left=5pt] {} (j15);

\path[] (i25) edge [thick, ->, in = 0, out=0, looseness=1, color=red]  node[near end, below left=5pt] {} (i26)
(i26) edge [thick, ->, in = 0, out=0, looseness=1, color=red]  node[near end, below left=5pt] {} (i210);

\path[] (j25) edge [thick, ->, in = 0, out=0, looseness=1, color=red]  node[near end, below left=5pt] {} (j26)
(j26) edge [thick, ->, in = 0, out=0, looseness=1, color=red]  node[near end, below left=5pt] {} (j210);
\end{tikzpicture}
\caption{}\label{fig:charge_network}
\end{subfigure}
    \caption{(a) Original charging network $\mathcal{G}$, (b) Charge-augmented network $\augnetworkflow$.}
\end{figure}
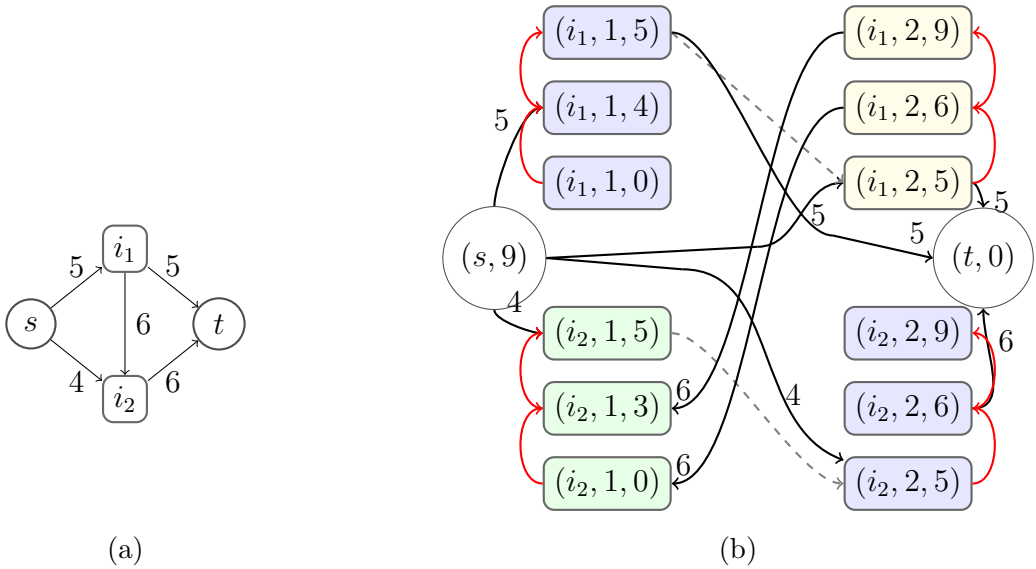
\end{example}
%\manxi{This example has an issue: s-j-t should be a path even though the battery is not depleted. Destination has 0 battery level only when charging is needed. This network has infinite capacity. Also, we reserved i for station, j for battery interval. Can we use $i_1$ and $i_2$ for stations?} \priya{Fixed}

If, for some $(s_k, t_k)$, there exists an $s_k$-$t_k$ path of length strictly less than $L$, an infinite amount of flow can be sent along this path. We assume that this is not the case, i.e., every $s_k$-$t_k$ path has a length of at least $L$, for all $k \in K$. For each $(s_k, t_k)$, we create the origin $(s_k, L)$ representing that a vehicle departs from $s_k$ with full battery, and $(t_k, 0)$ representing that a vehicle arrives at $t_k$ with no battery left. Note that, as we have assumed that all $s_k$-$t_k$ paths have length of at least $L$, if there is no $(s_k, L)$-$(t_k, 0)$ path in $\augnetworkflow$, a feasible solution does not exist. 

Furthermore, every path connecting $(s_k, L)$ and $(t_k, 0)$ corresponds to a charging strategy $(P, \chargeext)$ in $\Qext{}$. Recall from
Lemma \ref{lemma:finite} that we can restrict our attention to the EV flow vector that only sends positive flow along charging strategies $(P, \qhat) \in \Qext{}$. Therefore, the maximum EV flow vector of the charging network $\mathcal{G}$ can be equivalently computed as the maximum multicommodity flow of the charge-augmented network $\augnetworkflow$ with station capacity imposed on type I edges. We denote the edge load vector in $\augnetworkflow$ as $\hat{f} = (\hat{f}_{\ehat}^k)_{\ehat \in \Ehat, k \in K}$, where $\hat{f}_{\ehat}^k$ is the flow on edge $\ehat$ associated with $s_k$-$t_k$. The maximum multicommodity flow $\hat{f}^*$ of $\augnetworkflow$ can be computed as an optimal solution of a linear program. Then, we can construct the maximum EV flow vector $x^*$ using the flow decomposition theorem.

%Each extreme point charging strategy $(P, \hat q) \in \Qext{}$, $Pq$ is the corresponding path in $\augnetworkflow$. 
%Moreover, given flow $f$ in the charge-augmented network $\augnetworkflow$, we can recover an EV flow $(x_{P\hatq})_{(P, \hat q) \in \Qext{}}$ by defining $x_{P\hat q} = f_{\hat P}$, where $\hat P$ is the path corresponding to $(P, \hat q)$ in the charge-augmented network. Thus, we can turn the path-based formulation \eqref{P:max_finite} to an edge-based formulation with the edge load vector $\flow^k_{\ehat} = \sum_{(\hat P, \hat{q}) : \hat{P} \ni \ehat} \hat{x}^k_{(\hat P, \hat{q})}$ for all $\ehat \in \Ehat$. 

%$\hat{f}=(\hat(f)_{\ehat}^k)_{k \in K, \ehat \in \Ehat}$, where $\hat(f)_{\ehat}^k= \sum_{}\hat{x}_{\hat{P}}$  %\manxi{Change these to multi o-d. Assumptions should be added to the statement of theorems. Also, you need to explain why assumption 1 is needed for min cost flow. } 
{%$(\flow_{\ehat})_{\ehat \in \Ehat}$, where $\flow_{\ehat} = \sum_{\hat{q}\ni \ehat} \hat{x}_{\hat{q}}$ for all $\ehat \in \Ehat$. 
\begin{theorem}\label{theorem:max-flow}
    %Under Assumption \ref{as:infinite}, 
    The vector $\hat{f}^*$ induced by the maximum EV flow vector $x^*$ is the optimal solution of the following linear program: 
    \begin{subequations}\label{eq:maxflow:edgeformulation}
        \begin{align}
        \max_{\hat{f}} & ~ \sum_{k \in K} \sum_{\ehat \in \delta^{-}(t_k, 0)} \flow^k_{\ehat},\notag \\
    & ~ \sum_{k \in K} \sum_{\ehat \in \Eijone} \changebat_{\ehat} \flow^k_{\ehat} \leq \speed_{ij} z_{ij},  && \forall i \in I, \quad \forall j\in \{1, \dots, J\}, \label{subeq:constraint_one}\\
    &~ \sum_{j=1}^{J}z_{ij} = a_i, &&  \forall i \in I, \\
    & ~ \sum_{\ehat \in \delta^{+}(\vhat)} \flow^k_{\ehat} - \sum_{\ehat \in \delta^-(\vhat)} \flow^k_{\ehat} = 0,  && \forall \vhat\in \Vaug \setminus \{(s_k, L), (t_k, 0)\}, \, \forall k \in K, \\ 
    &~ \flow^k_{\ehat} \geq 0, && \forall \ehat \in \Ehat, \quad \forall k \in K,\\
&~ z_{ij} \geq 0, && \forall i \in I, \quad \forall j \in \{1, \dots, J\}, \label{subeq:constraint_last}
        \end{align}
    \end{subequations}
    where $\lambda_{\ehat}$ is the amount of battery charged when traversing the edge $\ehat \in \Eijone$. 
    %Moreover, any $\xhatopt$ that induces $\flowopt$ is a maximum flow vector, and $\xhatopt_{\qhat}$ is the volume of EV sent to take charging strategy $\qhat$.  
\end{theorem}
}

% \begin{align*}
% 
% \end{align*}Hence, to compute a maximum EV flow, it suffices to solve the edge-based max flow formulation in $\augnetworkflow$ with these additional capacity and resource allocation constraints. This immediately yields the following theorem. 
% From Lemma \ref{lemma:finite}, we know that we can restrict attention to flow vectors $\hat{x}= (\hat{x}_{\hat{q}})_{\hat{q} \in \hat{Q}}$. Then, the maximum EV flow problem can be equivalently posed as the maximum flow of the augmented network with capacity constraints posed for flows that use edges in each $\Eijone$ jointly as in \eqref{P:max_finite}. However, since the number of path in the augmented network is exponential in the number of nodes, we turn the path-based formulation \eqref{P:max_finite} in an edge-based formulation with the edge load vector $(\flow_{\ehat})_{\ehat \in \Ehat}$, where 
% \begin{align*}
% \flow_{\ehat} = \sum_{\hat{q}\ni \ehat} \hat{x}_{\hat{q}}, \quad \forall \ehat \in \Ehat. 
% \end{align*}
% Then, we further simplify linear program for computing the maximum flow problem as follows: 
%\priya{I think we don't need to explain the edge-path flow equivalence too much, since this is quite standard (and it should give us a bit of space)}\manxi{sure}

We now turn our attention to the minimum-cost EV flow problem, where a flow of value at least $D_k$ is routed from $s_k$ to $t_k$, for each $k \in K$, so as to minimize the sum of all monetary and time-costs \eqref{P:mincost}. In general settings, this problem is NP-hard. This is because in the special case with unlimited station capacity, the min-cost EV flow problem reduces to routing all the flow using a single EV optimal charging strategy, which has been shown to be NP-hard (Sec. \ref{sec:gasStation}). Under Assumption \ref{as:proportional}, we show that a minimum-cost EV flow can be computed in polynomial-time following similar analysis as for the maximum EV flow problem building on Lemma \ref{lem:polyChargeLevels}. %, we can obtain the following theorem, which is analogous to Theorem \ref{theorem:max-flow} for the minimum-cost setting. %, we can similarly compute the minimum-cost flow vector $\xmin$ using a polynomially-sized linear program as follows: 

\begin{theorem}\label{prop:min-cost} Under Assumption \ref{as:proportional}, the vector $\hat{f}^{\dagger}$ induced by the minimum-cost EV flow vector $\xmin$ is an optimal solution of the following linear program:  
    \begin{subequations}\label{eq:mincost:edgeformulation}
        \begin{align}
        \min_{\hat{f}} & \, \sum_{k \in K} \sum_{\ehat \in \Ehat} \gamma_{\ehat}\flow^k_{\ehat}, ~ s.t. ~ \text{$\hat{f}$ satisfies \eqref{subeq:constraint_one} -- \eqref{subeq:constraint_last}}, \sum_{\ehat \in \delta^-(t_k, 0)} \flow^k_{\ehat} \geq D_k, \,\, \forall k \in K, 
        \end{align}
    \end{subequations}
    where $\gamma_{\ehat}$ is the cost of traversing one unit of flow through the edge $\ehat \in \Ehat$, i.e., $\gamma_{\ehat} =( \tau_{ij} + (1+ \rho_{i})/r_{ij})\lambda_{\ehat}$ for all $\ehat \in \Eijone$ and $\gamma_{\ehat} =\ell_{ij, i'j'}$ for all $\ehat \in \Ehat_{ij,i'j'}$ and all $i, i' \in I, j, j' \in \{1, \dots J\}$.  
%    Moreover, any $\xhat^{\dagger}$ that induces $\hat{f}^{\dagger}$ is a maximum flow vector, and $\xhat^{\dagger}_{\qhat}$ is the volume of EV sent to take charging strategy $\qhat$.  
\end{theorem}

Thus, by solving the linear program given in Theorem \ref{theorem:max-flow} (resp. Theorem \ref{prop:min-cost}), we can compute an optimal solution to the maximum EV flow (resp. minimum-cost EV flow) problem in polynomial time. 
%    Based on Theorem \ref{theorem:max-flow} and Proposition \ref{prop:min-cost}, we conclude that through the graph augmentation approach, the maximum flow and the minimum-cost flow can both be computed as the optimal solution of finite linear programs with sizes in $|I|J$. 

\subsection{EV flow with edge capacities}\label{subsec:general}

In this section, we consider a generalized setting where apart from the station capacity constraints, each edge $e \in E$ also has a capacity $u_e \geq 0$. That is, an EV flow $x$ is feasible if it satisfies \eqref{eq:feasflow:cap} -- \eqref{eq:feasflow:zij}, and the edge-capacity constraints: 
\begin{align}\label{eq:edge_capacity}
    \sum_{(P, q) \in \Q : e \in P} x_{Pq} \leq u_e, \quad \forall e \in E.
\end{align} 
We show that the maximum EV flow and the minimum-cost EV flow problems with edge capacities are NP-hard, demonstrated by a reduction from \verb|PARTITION|, see Appendix \ref{apx:flow}.  

\begin{proposition}\label{thm:maxFlowHard}
	It is NP-hard to decide if there exists an EV flow $x$ that satisfies \eqref{eq:feasflow:cap} -- \eqref{eq:feasflow:zij} and \eqref{eq:edge_capacity} with $|x| \geq D$. 
\end{proposition}

Our fully polynomial-time approximation schemes build on the equivalence of the approximate separation and approximate optimization for the dual programs of \eqref{P:max} and \eqref{P:mincost} with edge capacity constraints \eqref{eq:edge_capacity}: 
  \begin{subequations}
\begin{align}
	\min & \sum_{i \in I} a_i y_i + \sum_{e \in E} u_e w_e,  \tag{$\mathrm{D}_{\mathrm{max-flow}}^{\mathrm{cap}}$} \label{D:max:cap} \\ 
	s.t. & \sum_{i \in I} \sum_{j = 1}^J \pi_{ij} q_{ij} + \sum_{e \in P} w_e \geq 1, && \forall (P, q) \in \Q, \label{constr:fptasdual1}\\
	& y_i \geq \pi_{ij} \speed_{ij}, && \forall i \in I, \quad \forall j \in \{ 1, \ldots, J\}, \\ 
	& \pi, w \geq 0,  
\end{align} 
\end{subequations}
where $w_e$ is the dual variable associated with constraint \eqref{eq:edge_capacity}, and  

\begin{subequations}
    \begin{align}
	\max~ & \sum_{k \in K} \phi_k D_k - \sum_{i \in I} a_i y_i - \sum_{e \in E} u_e w_e, \tag{$\mathrm{D}_{\mathrm{min-cost}}^{\mathrm{cap}}$}\label{D:mincost:cap} \\
s.t. ~ & \sum_{i \in I} \sum_{j = 1}^J (\tau_{ij} + \pi_{ij}) q_{ij} + \sum_{e \in P}( w_e + \ell_e) \geq \phi_k, ~  \forall (P, q) \in \Q_{s_kt_k}, \forall k \in K, \label{constr:fptasdual2} \\
 & y_i \geq \pi_{ij}r_{ij},  \quad \forall i \in I, \quad \forall j \in \{ 1, \ldots, J\}, \\ 
	& \pi, w, \phi \geq 0,  
\end{align} 
\end{subequations}
where $\phi_k$ is the dual variable associated with the demand constraint in \eqref{P:mincost}. 

 We will use the ellipsoid method to approximately solve \eqref{D:max:cap} (resp. \eqref{D:mincost:cap}). This requires a $(1 + \varepsilon)$-approximate separation oracle that returns a separating hyperplane for any candidate solution that violates \eqref{constr:fptasdual1} (resp. \eqref{constr:fptasdual2}) by a factor strictly larger than $(1 + \varepsilon)$. % $( w,  y,  \pi)$, which determines if a candidate solution is approximately feasible; if not, it returns a separating hyperplane.
 %In particular, this means that the oracle can approximately determinue whether or not \eqref{} and \eqref{} is violated by $\varepsilon$. 
To verify if \eqref{constr:fptasdual1} (resp. \eqref{constr:fptasdual2}) is approximately satisfied, it suffices to compute a $(1+\varepsilon)$-approximate min-cost single EV charging strategy $(P, q)$ where the edge cost is $w_e$ (resp. $w_e+\ell_e$) and the charge cost is $\pi_{ij}$ (resp. $\tau_{ij}+\pi_{ij}$) for every $e \in E$ and every $i \in I, j\in \{1, \dots, J\}$. This can be implemented in $O(poly(|I|J)/\varepsilon)$ time using the FPTAS by \cite{merting2015routing}. Following similar analysis as in Lemma \ref{lemma:finite}, we can equivalently reformulate the semi-infinite linear programs \eqref{D:max:cap} and \eqref{D:mincost:cap} as finite linear programs. This guarantees that the ellipsoid method can compute a $(1+\varepsilon)$-approximate dual solution in $O(poly(|I|J)/\varepsilon)$-time. Furthermore, by Lemma 6.5.15 in  \cite{grotschelGeometricAlgorithmsCombinatorial1988}, we can also obtain a $(1+ \varepsilon)$-approximate solution to the primal problems that compute the max EV flow and the min-cost EV flow in  $O(poly(|I|J)/\varepsilon)$-time. We present the algorithms in Appendix \ref{app:alg}.

\begin{proposition}
A $(1 + \varepsilon)$-approximate solution for the maximum EV flow problem and the min-cost EV flow problem with edge capacities can be computed in $O(poly(|I|J)/\varepsilon)$ time. %  There exists an FPTAS for the maximum EV flow with edge capacities problem, as well as for the minimum-cost EV flow problem with general costs and edge capacities.    
\end{proposition}

%A direct consequence of Theorem \ref{thm:maxFlowHard} is the problem of computing the {minimum-cost flow vector} with edge-capacity constraints is NP-hard. The minimum-cost flow problem (without edge-capacity constraints) is also NP-hard if additional assumptions (e.g. Assumption \ref{as:proportional}) are not made on the relationship between the edge distances and costs, as we have the constrained shortest path problem as a special case of the minimum-cost EV flow problem. 

%As done for the maximum flow problem, we can obtain an FPTAS for the minimum-cost flow problem (with general costs and edge capacities) by using Merting et al's algorithm and appealing to the equivalence of (approximate) separation and (approximate) optimization.

%\begin{theorem}\label{thm:mincostFPTAS}
%	If there are no flow-regenerating cycles in the network, there exists an FPTAS for the problem of computing a minimum-cost flow respecting edge capacities (without Assumption \ref{as:proportional}).  
%\end{theorem}

\color{black}

%, or the relationship between the costs and distances of edges does not satisfy Assumption \ref{assumption1}. In such settings, the problem of computing a maximum flow vector or minimum 

\bibliographystyle{splncs04}
\bibliography{references}




% LNCS Homepage, \url{http://www.springer.com/lncs}. Last accessed 4
% \end{thebibliography}
\appendix
\section{Proofs from Section \ref{sec:gasStation}}

\noindent\emph{Proof of Lemma \ref{lem:polyChargeLevels}.} We first prove that any optimal charging strategy $(P^*, \qopt)$ must satisfy \eqref{eq:charge_opt_condition}. Let $P= s_k\, (i_1, j_1)\, \cdots \, (i_m, j_m) \,t_k$; for ease of exposition, we define $v_n = (i_n, j_n)$. 
Under the condition that the amount of battery charge at every visited node on $\pathopt$ is positive, we know that $\qopt_{v_i} \in [\minbat{v_{i+1}} + d_{v_i, v_{i+1}} - \batteryin_{v_i},  \maxbat{v_i} - \batteryin_{v_i}]$ for all $v_i \in \pathopt$. Here, $\qopt_{v_i} = \minbat{v_{i+1}} + d_{v_i, v_{i+1}} - \batteryin_{v_i}$ is the minimum charge that the vehicle needs to be eligible charging in the next visited node $v_{i+1}$, and $\qopt_{v_i} = \maxbat{v_i} - \batteryin_{v_i}$ is the maximum charge that the vehicle can get from node $v_i$. 

Suppose that when $c_{v_i} > c_{v_{i+1}}$, we have $\qopt_{v_i}> \minbat{v_{i+1}} + d_{v_i, v_{i+1}} - \batteryin_{v_i}$. Then, we consider another charging strategy $q'$ such that $q'_{v_j} = \qopt_{v_j}$ for any $j \neq \{i, i+1\}$, $q'_{v_i} = \qopt_{v_i} -\epsilon$ and $q'_{v_{i+1}} = \qopt_{v_{i+1}} + \epsilon$ for sufficiently small $\epsilon>0$. Such a $q'$ is feasible, and reduces the cost of charging by $(c_{v_i} - c_{v_{i+1}}) \epsilon>0$, hence contradicting the fact that $\qopt$ is optimal. Therefore,  $\qopt_{v_i}= \minbat{v_{i+1}} + d_{v_i, v_{i+1}} - \batteryin_{v_i}$ when $c_{v_i} > c_{v_{i+1}}$. Analogously, when $c_{v_i} \leq c_{v_{i+1}}$ but $\qopt_{v_i} <\maxbat{v_i} - \batteryin_{v_i}$, we can increase $\qopt_{v_i}$ by $\epsilon$ and decrease $\qopt_{v_{i+1}}$ by $\epsilon$ for some $\epsilon \in (0, \maxbat{v_i} - \batteryin_{v_i}- \qopt_{v_i}]$; this strictly reduces the cost, contradicting the optimality of $(P^*, q^*)$. Thus, $\qopt_{v_i} =\maxbat{v_i} - \batteryin_{v_i}$.

Equation \eqref{eq:charge_opt_condition} indicates that the battery level of vehicles when arriving at any node $v \in V$ given the optimal charging strategy must be either at the minimum allowable level $\bminv$ or at the level of $\maxbat{v'}- d_{v'v}$, i.e., a vehicle leaving from $v'$ with maximum possible battery level and consume $d_{v'v}$ amount of battery for traversing from $v'$ to $v$. Here, $v'$ can be any neighbor of node $v$ such that the battery level $\maxbat{v'}- d_{v'v}$ is within the feasible interval of node $v$. \hfill $\square$

\section{Proofs from Section \ref{sec:maxFlow}}\label{apx:flow}

\noindent \emph{Proof of Lemma \ref{lemma:finite}}. We note that that constraint \eqref{constr:dualsep} is equivalent to \[ \min_{(P, q) \in \Q} \sum_{i \in I} \sum_{j = 1}^J \pi_{ij} q_{ij} \geq 1.\] 
Since $\Q = \cup_{P \in \mathcal{P}} \{(P, q) : q \in Q_P\}$, 
 \[ \min_{(P, q) \in \Q} \sum_{i \in I} \sum_{j = 1}^J \pi_{ij} q_{ij} =  \min_{(P, \hat q) \in \Qext{}} \sum_{i \in I} \sum_{j = 1}^J \pi_{ij} \hat q_{ij}, \]
 where $\hat\Q = \cup_{P \in \mathcal{P}} \{(P, \hat q) : \hat q \in \hat{Q}_P\}$, and $\hat{Q}_P$ is the set of extreme points of $Q_P$.  Thus, constraint \eqref{constr:dualsep} is equivalent to $\sum_{i \in I} \sum_{j = 1}^J \pi_{ij} \hat q_{ij} \geq 1$ for all $(P, \hat q) \in \Qext{}$; it immediately follows that \eqref{D:max} is equivalent to \eqref{D:max_finite}. By strong duality, we conclude that \eqref{P:max} is equivalent to \eqref{P:max_finite}. \hfill $\square$. \\[5pt]

\noindent \emph{Proof of Theorem \ref{prop:min-cost}}. The minimum-cost EV flow problem can be formulated as the following (semi-infinite) linear program 
 \begin{subequations}
     \begin{align}
         \min_{x,z}  ~ &\sum_{(P, q) \in \Q} \left( \sum_{i \in I} \sum_{j=1}^J (\tau_{ij} + (1 + \rho_i)/\speed_{ij})q_{ij} +  \sum_{e \in P} \ell_e \right) x_{P, q}, \tag{$\mathrm{P_{min-cost}}$}\\
         \text{s.t.} ~& \sum_{(P, q) \in \Q} q_{ij} x_{P,q} \leq \speed_{ij} z_{ij}, \qquad \forall i \in I, \quad \forall j \in \{1, \ldots, J\}, \\ 
         & \sum_{(P, q) \in \Q_{k}} x_{Pq} \geq D_k, \qquad  \qquad \forall k \in K, \\ 
     ~~&~~ \sum_{j=1}^{J} z_{ij} = \nslot_i, \qquad \qquad \quad ~~~~ \forall i \in I,  \\
    ~~&~~ z_{ij} \geq 0,  ~\qquad  \qquad \qquad \qquad \forall i \in I, ~ \forall j \in \{1, \ldots, J\}. 
     \end{align}
 \end{subequations}
 Its dual is 
 \begin{subequations}
    \begin{align}
	\max~ & \sum_{k \in K} \phi_k D_k - \sum_{i \in I} a_i y_i, \notag \\
s.t. ~ & \sum_{i \in I} \sum_{j = 1}^J \left(\left(\tau_{ij} + \frac{1 + \rho_i}{\speed_{ij}}\right) + \pi_{ij}\right) q_{ij} + \sum_{e \in P} \ell_e \geq \phi_k,\quad \forall (P, q) \in \Q_{k}, \forall k \in K, \\
 & y_i \geq \pi_{ij}\speed_{ij}, \quad \forall i \in I, \quad  \forall j \in \{1, \ldots, J\}, \\ 
	& \pi, w \geq 0.  
\end{align} 
\end{subequations}
We note that $\min_{(P, q) \in \Q} \sum_{i \in I} \sum_{j = 1}^J (\tau_{ij} + \frac{1 + \rho_i}{\speed_{ij}}+ \pi_{ij}) q_{ij} + \sum_{e \in P} \ell_e \geq \phi_k$ if and only if \[\min_{(P, \hat q) \in \Qext{k}} \sum_{i \in I} \sum_{j = 1}^J (\tau_{ij} + \frac{1 + \rho_i}{\speed_{ij}} + \pi_{ij}) \hat q_{ij} + \sum_{e \in P} \ell_e \geq \phi_k,\] where $\Qext{k} =   \cup_{P \in \mathcal{P}_k} \{(P, \hat q) : \hat q \in \hat{Q}_P\} $. By strong duality, there exists an optimal solution $(\xmin, z^\dagger)$ such that the support of $\xmin$ is a subset of $\Qext{}$. %Hence, by the same argument as before, there exists an optimal solution in which every vehicle's charge level upon arrival and departure from a node $(i, j)$ belongs to $\Batteryin_{ij} \cup \Batteryout_{ij}$. , 
Furthermore, each charging strategy in $\Qext{}$ corresponds to an $(s_k, L)-(t_k, 0)$ path in $\augnetworkflow$. We define edge costs to be 0 for type II edges, $\ell_{ii'}$ for type III edges, and  $(\tau_{ij} + (1 + \rho_i)/\speed_{ij})\lambda_e$ for type I (charging) edges where $\lambda_e$-units of charge are obtained. Then, for any charging strategy $(P, \hat q) \in \Qext{}$, the cost of the corresponding path in the charge-augmented network is $ \sum_{i \in I} \sum_{j=1}^J (\tau_{ij} + (1 + \rho_i)/\speed_{ij})\hat q_{ij} +  \sum_{e \in P} \ell_e$. 

The minimum-cost EV flow vector of the charging network $\mathcal{G}$ can equivalently be computed as the minimum-cost multicommodity flow of the charge-augmented network $\augnetworkflow$ with station capacity constraints imposed on type I edges. Recall that we denote the edge load vector in $\augnetworkflow$ as $\hat{f} = (\hat{f}_{\ehat}^k)_{\ehat \in \Ehat, k \in K}$, where $\hat{f}_{\ehat}^k$ is the flow on edge $\ehat$ associated with $s_k$-$t_k$. The minimum-cost flow $\hat{f}^\dagger$ of $\augnetworkflow$ can be computed as an optimal solution of the linear program \eqref{eq:mincost:edgeformulation}. Then, we can construct the minimum-cost EV flow vector $\xmin$ using the flow decomposition theorem.  \hfill $\square$ \\

\noindent \emph{Proof of Proposition \ref{thm:maxFlowHard}}. We prove this via a reduction (similar to the one given in \cite{hall2007multicommodity}) from the \verb|PARTITION| problem: Given $n$ integer numbers $z_1, \ldots, z_n \in \mathbb{N}$ with $\sum_{i = 1}^n z_i = 2L$, one must decide whether there exists $B \subseteq \{1, \ldots, n\}$ such that $\sum_{i \in B} z_i = L$. \\
	
	Consider the following network. 
 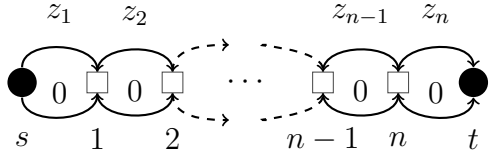
\begin{figure}[ht]
     \centering
     \begin{tikzpicture}[
station/.style={rectangle, draw=black!70,fill=white, minimum size=1mm}
]
%Nodes
\node[fill=black, style=circle]      (s) at (0, 0) {};
\node[] () at (0, -0.75) {$s$};
\node[station]      (z1) at (1, 0) {};
\node[] () at (1, -0.75) {$1$};
\node[station] (z2) at (2, 0) {}; 
\node[] () at (2, -.75) {$2$}; 
\node[] () at (3, 0) {$\cdots$}; 
\node[station] (zn1) at (4, 0) {}; 
\node[] () at (4, -.75) {$n-1$}; 
\node[station] (zn) at (5, 0) {}; 
\node[] () at (5, -.75) {$n$}; 
\node[fill=black, style=circle]      (t) at (6, 0) {};
\node[] () at (6, -0.75) {$t$};
%Lines
\path[] (s) edge [thick, ->, out = 90, in=90, looseness=1, ]  node[midway, above=5pt] {$z_1$} (z1);
\path[] (s) edge [thick, ->, out = -90, in=-90, looseness=1, ]  node[midway, above=1pt] {$0$} (z1);

\path[] (z1) edge [thick, ->, out = 90, in=90, looseness=1, ]  node[midway, above=5pt] {$z_2$} (z2);
\path[] (z1) edge [thick, ->, out = -90, in=-90, looseness=1, ]  node[midway, above=1pt] {$0$} (z2);

\path[] (z2) edge [thick, ->, dashed, out = 90, in=180, looseness=1, ]  node[midway, above=5pt] {} (2.75, .5);
\path[] (z2) edge [thick, ->, dashed, out = -90, in=180, looseness=1, ]  node[midway, above=1pt] {} (2.75, -.5);

\path[] (3.25, .5) edge [thick, ->, dashed, in = 90, out=180, looseness=1, ]  node[midway, above=5pt] {} (zn1);
\path[] (3.25, -.5) edge [thick, ->, dashed, in = -90, out=180, looseness=1, ]  node[midway, above=1pt] {} (zn1);

\path[] (zn1) edge [thick, ->, out = 90, in=90, looseness=1, ]  node[midway, above=5pt] {$z_{n-1}$} (zn);
\path[] (zn1) edge [thick, ->, out = -90, in=-90, looseness=1, ]  node[midway, above=1pt] {$0$} (zn);

\path[] (zn) edge [thick, ->, out = 90, in=90, looseness=1, ]  node[midway, above=5pt] {$z_{n}$} (t);
\path[] (zn) edge [thick, ->, out = -90, in=-90, looseness=1, ]  node[midway, above=1pt] {$0$} (t);
% \draw  (c) edge[bend right=300,auto=right]  node [midway, above, fill=none]{} (H);
% \draw  (c) edge[bend right=60,auto=right]  node [midway, below, fill=none]{} (H);
\end{tikzpicture}
     \caption{Network for PARTITION instance}
     \label{fig:enter-label}
 \end{figure}
In this network, each station $i$ has a speed of $\speed_i = 0$; the top  $e_{i-1, i}$ arc has a distance of $z_i$, while the bottom $e'_{i-1, i}$ arc has a distance of 0. Every arc has a capacity of 1. 
	 
	 If there exists $B$ such that $\sum_{i \in B} z_i = L$, we define $P_1$ to be the $s$-$t$ path taking the top $e_{i-1, i}$ arc for all $i \in B$, and the bottom $e'_{i-1, i}$ arc for all $i \notin B$. $P_2$ is the path $G \setminus P_1$. Since the length of $P_1$ and $P_2$ is exactly $L$, taking path $P_1$ or $P_2$ without charging at any node is a feasible strategy. Thus, there exists an EV flow of value 2. % zero charging amount at each node so $(P_1, 0)$ and $(P_2, 0)$ are feasible charging strategies. Hence $x_{P_1,0} = 1 = x_{P_2,0}$ is a feasible solution. 

	 We now prove the other direction. Suppose there exists a flow $x^*$ of value 2. Since all arcs have a capacity of 1, $\sum_{(P, q) \in \Q : e \in P} x^*_{P,q} = 1$ for all $e \in E$. Furthermore, as charging stations have a speed of 0, the support of $x^*$ consists of charging strategies $(P, q)$ such that $q_v=0$ for all $v \in V$ and $\sum_{e \in P} d_e \leq L$. Note that the average path length is $\sum_{(P, q) \in \Q} x_{P,q} \cdot (\sum_{e \in P} d_e) = \sum_{e \in E} d_e \cdot 1 = \sum_{i = 1}^n z_i = 2L$.  
	  Suppose for the sake of contradiction that there exists $(P', q') \in supp(x^*)$ with $\sum_{e \in P'} d_e < L$. Then 
	 $\sum_{(P, q) \in \Q} x^*_{P,q} \cdot \sum_{e \in P} d_e =  (2 - x^*_{P',q'})L  + x^*_{P',q'}\sum_{e \in P'} d_e  < 2L$, which is a contradiction. So, every path in the support of $x^*$ must have a length of exactly $L$, and hence induces a subset $B$ with $\sum_{i \in B} z_i = L$. 
  
  Thus, the NP-hard problem \verb|PARTITION| reduces to the problem of deciding if there exists an EV flow of value at least 2 in the above network. \hfill $\square$\\

\subsection{FPTAS for maximum EV flow and minimum-cost EV flow problems with edge capacities}\label{app:alg}

\begin{algorithm}[ht] \label{alg:max-fptas} \caption{FPTAS for maximum EV flow with edge capacities.}
	\SetAlgoLined	
	\vspace{5pt}
	Using the ellipsoid method  (with the FPTAS algorithm in \cite{merting2015routing} as the separation oracle) solve \eqref{D:max:cap}; let $(P^{(1)}, q^{(1)}), \ldots, (P^{(N)}, q^{(N)})$ be the charging strategies corresponding to the separating hyperplanes produced by the ellipsoid method. \\ 
\Return{} the optimal solution of the following linear program:
\begin{align*}
    \max &~ \sum_{m=1}^N  x_{P^{(m)}q^{(m)}}, \notag  \\
    \text{s.t.}&~ \sum_{m=1}^N q^{(m)}_{ij} x_{P^{(m)}q^{(m)}} \leq \speed_{ij} z_{ij}, && \forall i \in I, \quad \forall j \in \{ 1, \ldots, \maxthresh\},\\
    & ~ \sum_{j=1}^\maxthresh z_{ij} = \nslot_i, && \forall i \in I,  \\ 
    & \sum_{m : e \in P^{(m)}} x_{P^{(m)}q^{(m)}} \leq u_e, && \forall e \in E, \\ 
    &~ x_{P^{(m)}q^{(m)}} \geq 0, && \forall m  \in \{1, \ldots, N\}.
\end{align*}
\end{algorithm}

\begin{algorithm}[ht] \label{alg:min-fptas} \caption{FPTAS for minimum-cost EV flow with edge capacities. }
	\SetAlgoLined	
	\vspace{5pt}
	Using the ellipsoid method (with the FPTAS algorithm in \cite{merting2015routing} as the separation oracle), solve \eqref{D:mincost:cap}. Let $(P^{(1)}, q^{(1)}), \ldots, (P^{(N)}, q^{(N)})$ be the charging strategies corresponding to the separating hyperplanes generated by the ellipsoid method. \\ 
\Return{} the optimal solution of the following linear program: 
\begin{align*}
    \min &~ \sum_{m=1}^N  \left( \sum_{i \in I} \sum_{j=1}^J (\tau_{ij} + (1 + \rho_i)/\speed_{ij})q^{(m)}_{ij} +  \sum_{e \in P^{(m)}} \ell_e \right)  x_{P^{(m)}q^{(m)}} ,\notag  \\
    \text{s.t.}&~ \sum_{m=1}^N q^{(m)}_{ij} x_{P^{(m)}q^{(m)}} \leq \speed_{ij} z_{ij}, ~\quad  \forall i \in I,  \quad \forall j \in \{ 1, \ldots, \maxthresh\},\\
    &  \sum_{j=1}^\maxthresh z_{ij} = \nslot_i, \qquad \forall i \in I, \\ 
    & \sum_{m : P^{(m)} \in \mathcal{P}_{k}} x_{P^{(m)}q^{(m)}} \geq D_k, \qquad \forall k \in K, \\ 
    & \sum_{m : e \in P^{(m)}} x_{P^{(m)}q^{(m)}} \leq u_e, \qquad \quad \forall e \in E, \\ 
    &~ x_{P^{(m)}q^{(m)}} \geq 0, \qquad \qquad \quad \qquad \forall m \in \{ 1, \ldots, N\}.
\end{align*}
\end{algorithm}

\end{document}